\documentclass[preprint,12pt]{elsarticle}

\usepackage{amssymb}
\usepackage{amsmath}
\usepackage{graphicx}
\usepackage{subfigure}
\usepackage{caption}
\usepackage{makecell}
\usepackage{booktabs}
\usepackage{algorithm}
\usepackage{algpseudocode}
\usepackage{color}

\usepackage{xspace}
\newcommand{\tool}{TLSAN\xspace}

\journal{Neurocomputing}

\begin{document}
\begin{sloppypar}

\begin{frontmatter}



\title{TLSAN: Time-aware Long- and Short-term Attention Network for Next-item Recommendation}

\author[a]{Jianqing Zhang}
\ead{jqzy4869@gmail.com}
\author[a]{Dongjing Wang\corref{cor1}}
\cortext[cor1]{Corresponding author.}
\ead{dongjing.wang@hdu.edu.cn}
\author[a]{Dongjin Yu}
\ead{yudj@hdu.edu.cn}
\address[a]{School of Computer Science and Technology, Hangzhou Dianzi University, China}

%

\begin{abstract}
	Recently, deep neural networks are widely applied in recommender systems for their effectiveness in capturing/modeling users' preferences. Especially, the attention mechanism in deep learning enables recommender systems to incorporate various features in an adaptive way. Specifically, as for the next item recommendation task, we have the following three observations: 1) users' sequential behavior records aggregate at time positions (``time-aggregation''), 2) users have personalized taste that is related to the ``time-aggregation'' phenomenon (``personalized time-aggregation''), and 3) users' short-term interests play an important role in the next item prediction/recommendation. In this paper, we propose a new \underline{T}ime-aware \underline{L}ong- and \underline{S}hort-term \underline{A}ttention \underline{N}etwork (\tool) to address those observations mentioned above. Specifically, \tool consists of two main components. Firstly, \tool models ``personalized time-aggregation'' and learn user-specific temporal taste via trainable personalized time position embeddings with category-aware correlations in long-term behaviors. Secondly, long- and short-term feature-wise attention layers are proposed to effectively capture users' long- and short-term preferences for accurate recommendation. Especially, the attention mechanism enables \tool to utilize users' preferences in an adaptive way, and its usage in long- and short-term layers enhances \tool's ability of dealing with sparse interaction data. Extensive experiments are conducted on Amazon datasets from different fields (also with different size), and the results show that \tool outperforms state-of-the-art baselines in both capturing users' preferences and performing time-sensitive next-item recommendation.
\end{abstract}



\end{frontmatter}


\section{Introduction}

As people enjoy the convenience of the Internet every day, such as shopping online, massive amounts of data have been generated and recorded. Especially, those data may indicate important information, such as users' preferences and behavior patterns, which can be utilized by recommender systems \cite{rendle2009bpr, wu2017recurrent} to provide personalized services or contents for users and promote their experience. 

Existing recommender systems are based on various methods and strategies, such as collaborative filtering (CF)~\cite{chen2020deep, qin2020sequential}, Markov chains~\cite{rendle2010factorizing, he2016fusing}, matrix factorization~\cite{chen2020deep, khan2020enriching}, Bayesian probabilistic models~\cite{rendle2009bpr, morise2019bayesian} and deep neural networks (DNN)~\cite{he2016vbpr, zhou2018personalized}. In recent years, attention mechanism~\cite{vaswani2017attention, cao2020position} begins to be used in many tasks, including recommendation, to incorporate various features adaptively and increase model training speed. Especially, attention can be combined with various methods without adding too much model complexity~\cite{cheng2018neural}. Furthermore, next-item recommender systems model users' records and corresponding temporal context as behavior sequences to utilize the context in a better way. Specifically, users' preferences can be captured from the entire sequence, and it can help make accurate predictions to provide personalized services.

\begin{figure}
	\centering
	\includegraphics[scale=0.66]{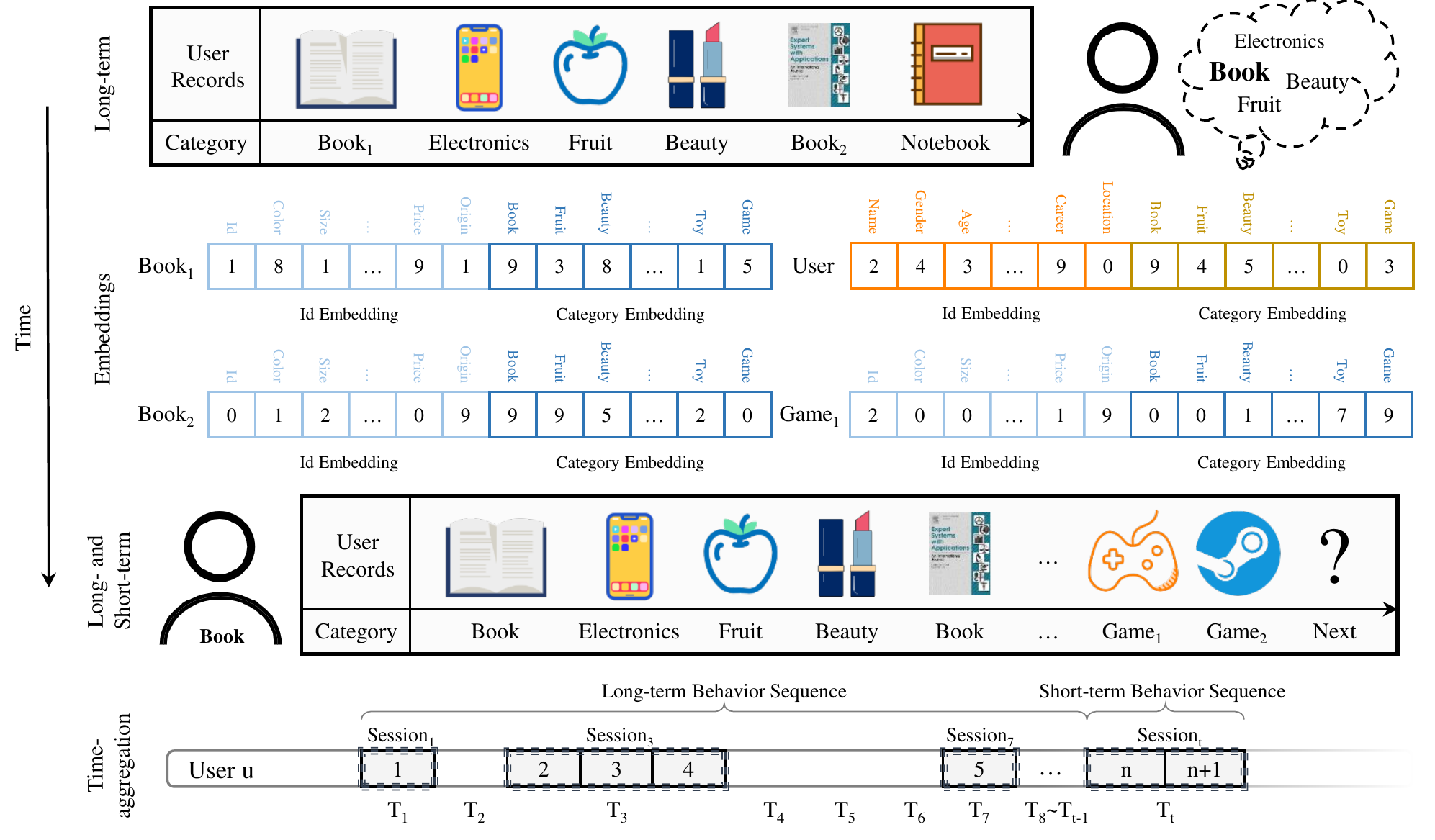}
	\caption{The user's long-term preferences can be inferred from her/his long-term behaviors, and the user's short-term interests are reflected in her/his short-term behaviors, both of which change over time and can be captured by the user's embedding. Specifically, $Book_1$ and $Book_2$ are in long-term behaviors while $Game_1$ is in short-term behaviors. (1) \textbf{Long-term}: If the prominent item category in user's behavior records is book, her/his category will be book. (2) \textbf{Embeddings}: We concatenate ID embedding and category embedding as user embedding. Each dimension of the embedding is manually labeled for convenience of description. (3) \textbf{Long- and Short-term}: Considering only long-term preferences of users without strengthening the impact of recent interests can lead to inaccurate recommendations. (4) \textbf{Time-aggregation}: Behavior records generally aggregate at time positions, which may be different for each user who has personalized behavior patterns.}\label{fig:1}
\end{figure}

However, existing approaches still face five main issues: (1) \textbf{Category-aware correlations modeling}: Most existing approaches implicitly group users, such as user-based CF~\cite{chen2020deep, koohi2016user, bellogin2012using} and social circle~\cite{purushotham2012collaborative, qian2013personalized}, which can not explicitly model the correlation between item category. Besides, some works~\cite{papagelis2005qualitative, yang2012circle} explicitly incorporate user category information as users' feature at the cost of high time complexity. Furthermore, users' categories may change dynamically over time and it is not accurate to model user as a particular category at each time position. (2) \textbf{Feature-level representation learning}: As is shown in Figure \ref{fig:1} (Embeddings), each dimension in the item's embedding have specific effects, which may make different contribution to the tasks of prediction/recommendation. For example, the user in Figure \ref{fig:1} is more interested in books ($Book_1$ and $Book_2$) than game ($Game_1$), so the weight of the feature that represents book is 9. Furthermore, the user prefers $Book_1$ to $Book_2$, which should be incorporated when we calculate her/his preference score. (3) \textbf{Personalized time-aggregation}: Each user's behavior records generally aggregate at time positions (``time-aggregation''), as shown in Figure \ref{fig:1} (Time-aggregation layer). Furthermore, each person has her/his own personalized behavior patterns, so different people react differently to similar contexts, such as interval of ``time-aggregation''. We call this phenomenon as ``personalized time-aggregation'', which is also consistent with our daily behavior patterns. For example, some people who are not active on normal days may generate a lot of shopping records on Amazon on some certain days like Black Friday. Besides, some active users do not enjoy the fun of Black Friday because of the complex coupons. Therefore, the recommendation mechanism for two kinds of users are different since they have personalized behavior patterns and ``time-aggregation'' phenomenon.  (4) \textbf{Long- and short-term preferences integration}: The next recommendation task is \textbf {time-sensitive}, and users' recent behaviors (short-term) generally play a more important role than users' actions/records that happened long time ago (long-term) . As shown in Figure \ref{fig:1} (long- and short-term), the user like books, and he/she become interested in games recently. Therefore, both users' long- and short-term preferences are important for accurate prediction and recommendation, although they may have different influences.  (5) \textbf{Data sparsity problem}: The data in many real-world scenarios is quite sparse, since the amount of users or items may reach millions easily. Especially, it is difficult to capture users' current preferences accurately only from sparse behaviors data.

Some recommendation methods are proposed to solve one or several of the issues mentioned above. For example, ATRank~\cite{zhou2018atrank} captures users' long-term preferences by self-attention and models the correlation between target item and the historical items by vanilla attention, but it ignores users' individual information, such as their category and temporal preferences. CSAN~\cite{huang2018csan} considers different features by feature-wise self-attention and describes ``time-aggregation'' with position matrix. However, the position matrix is untrainable, which may fail to capture personalized position preferences and cause an order of magnitude difference between position matrix and embeddings. SHAN~\cite{ying2018sequential} adopt a hierarchical attention network to capture users' long- and short-term preferences separately, but the user embedding in SHAN does not incorporate users' information, such as category. Besides, SHAN ignores time decay in long-term layer. In PACA~\cite{cao2020position}, the trainable global position embedding for modeling the effect of records in each position is the same for all the users, which may influence its ability of generalize personalized results. However, to the best of our knowledge, there is no attempt to address all those four issues.

In this paper, we propose a new recommendation model named \textbf{\underline{T}ime-aware \underline{L}ong- and \underline{S}hort-term \underline{A}ttention \underline{N}etwork (\tool)} to address those issues mentioned above.
Especially, \tool can effectively incorporate features of users and items as well as their correlations. Besides, the devised attention mechanism enables \tool to capture users' long- and short-term preferences from their behavior sequences and leverage both preferences adaptively for accurate prediction/recommendation. 
Specifically, the proposed approach \tool consists of two main parts: 
(1) \textbf{Personalized time position embedding module} is designed to model users' ``personalized time-aggregation'' and capture user-specific temporal preferences. Especially, behavior records at different time positions may have different contributions to the prediction/recommendation tasks, which depends on users' personalized behavior patterns.
(2) \textbf{Long- and short-term feature-wise attention layers} use long-term layer to capture users' long-term preferences and adopt a short-term layer to emphasize users' short-term interests. In sparse dataset, short-term interests can be the main determinants of predicting the next item and alleviating the \textbf{data sparsity problem} due to its small data demands. Specifically, feature-wise attention (a more fine-grained attention) can help to effectively capture users' long and short-term preferences in each dimension and leverage both preferences for better recommendation. Besides, we model the category-aware interactions between features of users or items in user-user or user-item pairs by dynamic user category extraction before model training, which does not increase model complexity. What is more, multi-heads integration is adopt to model the information from different semantic sub-spaces and integrate them as well as their correlations in a unified and parallel way to further improve the performance of recommendation.

Previous models utilized deep neural networks to incorporate a lot of factors at the cost of more pre-training time and higher model complexity. \tool utilizes attention mechanism comprehensively and adds more parallelism to itself, so that \tool can converge quickly and achieve good results. Due to the lightness of attention mechanism as well as its concise principle, \tool can be easily adjusted to various datasets and scenarios in an efficient way.

The main contributions in this paper are as follows:
\begin{itemize}
	\item We propose a \textbf{\underline{T}ime-aware \underline{L}ong- and \underline{S}hort-term \underline{A}ttention \underline{N}etwork (\tool)} to capture users' temporal preferences from their historical behavior sequences for accurate next item recommendation.
	\item To strengthen time-sensitive next-item recommendation ability and deal with sparsity, we firstly capture weakened users' long-term preferences by feature-wise attention in long-term layer with the personalized time position embeddings dynamically weakening long-term behaviors. Then we emphasize the effect of users' short-term interests by feature-wise attention in short-term layer. 
	\item In the above two layers, we incorporate category-aware user-user and user-item correlations without increasing model complexity by dynamic user category extraction. Besides, multi-head strategy is introduced to improve parallelism and find correlations between attention blocks. 
	\item Experiments on several kinds of Amazon public datasets with different size show that \tool outperforms state-of-the-art baselines and can maintain effectiveness on different recommendation scenarios.
\end{itemize}

The remaining of this paper is organized as follows. Section~\ref{sec_related_work} describes existing works related to our approach \tool. Section~\ref{sec_proposed_model} introduces \tool in details, including the framework and the main components. In Section~\ref{sec_experiments}, we evaluate \tool on several kinds of datasets with different size, and also explore the impact of hyper-parameters as well as \tool's key components. Finally, the conclusion and future works are provided in Section~\ref{sec_conclusion_future}.

\section{Related Work}
\label{sec_related_work}
Earlier next-item and sequence recommendation utilized traditional methods, such as collaborative filtering and Markov Chain. Later, the emergence of neural networks further enhanced recommender systems' ability of extracting users' preferences. However, most deep neural networks based methods are generally more complex, which influences their parallelism and interpretability, especially when many factors are incorporated. The attention mechanism that is widely applied in natural language processing (NLP) and Computer vision (CV), has shown its effectiveness in recommender systems. The attention mechanism has more parallelism than convolutional neural networks (CNN) and recurrent neural networks (RNN), and provides more possibilities of constructing powerful networks. Besides, researchers have proposed various methods and strategies, such as position embedding, to capture time series information for better sequential recommendation.

\subsection{Sequential Recommendation}

The methods used in existing sequential recommender systems can be classified into three main categories: collaborative filtering (CF), Markov chain (MC) and neural networks.

CF based methods: Earlier, with the assumption that users who have made similar behaviors may behave similarly next time, researchers utilize CF~\cite{chen2020deep, qin2020sequential, he2016ups} to perform next item recommendation. However, this kind of recommender systems are mainly based on statistical method~\cite{ungar1998formal}, which cannot accurately distinguish similar users.

MC based methods: To capture the sequential information in the sequence, researchers also introduce MC to the recommendation models~\cite{rendle2010factorizing, he2016fusing}. These methods can well model sequential patterns (short-term interests) with context information as Fossil~\cite{he2016fusing}, and are also good at predicting the next item. However, MC only considers the correlations between adjacent items~\cite{rendle2010factorizing}, so the models based on MC may have difficulty in capturing the users' long-term preferences. Besides, such methods cannot effectively model the dynamic changes of original context over time. 

Neural networks based methods: Multi-layer perceptions (MLPs) can introduce non-linear user-item correlations~\cite{he2017neural}, but it is not easy to determine hyper-parameters, such as how many layers are sufficient for the tasks. CNN can be utilized to extract features from texts, audios and pictures\cite{he2016vbpr, van2013deep, hao2019real}. For example, Realtime-MF~\cite{hao2019real} employs a CNN to obtain event embedding from the related words in reviews and the researchers find that the frequent words can describe the event well, but it is not easy to capture semantics without enough texts. RNN can capture dynamic time series information~\cite{zhou2018personalized, wang2019cross, wu2017recurrent}. As important variants of RNN, LSTM~\cite{jozefowicz2015empirical} and GRU~\cite{luo2019adaptive} use the gating mechanism to weaken the impact of long-term preferences, but they need much time to train the effective models. To build representations of both past and future contexts, recent approaches also apply bi-directional encoders~\cite{bansal2016ask}. Recent recurrent recommendation model CDHRM devises a cross-domain user-level RNN to capture global users' dynamic preferences and two domain-specific session-level RNNs to separately capture users' specific preferences in different domains, and fuse these two kinds of RNN to obtain comprehensive users' preferences~\cite{wang2019cross}. However, the concatenation and product fusion method in CDHRM is still not adaptive enough to fuse domain-specific RNNs together from different domains with different preferences, behavior distribution, etc.

\subsection{Attention Mechanism}

Attention mechanism is similar to human's visual attention: we always focus on the most important part of what we see. It has been widely utilized in many other fields, such as NLP~\cite{shen2018disan, yin2016abcnn} and CV~\cite{thanikasalam2019target, zhang2020multimodal}. Specifically, attention mechanism makes it easy to memorize various remote dependencies or focus on important parts of the input. In addition, attention-based methods are often more interpretable~\cite{sha2017interpretable} than traditional deep learning models.

Parallelism of attention mechanism and its preferences that can be fused with other models are widely applied in various kinds of task. For example, Transformer~\cite{vaswani2017attention} has been widely utilized in NLP, CV, recommender systems and other fields since proposed~\cite{ni2019modeling, cao2019cross}. It fully takes advantage of attention mechanism, and further improves parallelism with multi-heads~\cite{lyu2019attention}. ATRank~\cite{zhou2018atrank} applies multi-head self-attention components in Transformer to recommendation tasks, speeding up training and improving prediction ability, but it does not introduce users' individual information. CSAN~\cite{huang2018csan} introduces a feature-wise self-attention mechanism with bi-directional position embedding on input sequence to better discover internal correlations within sequences, which need lots of trainable parameters increasing model complexity. SHAN~\cite{ying2018sequential} utilizes original attention mechanism to capture users' long- and short-term preferences and introduce context information by user embedding, but it ignores time decay in the long-term behaviors and the user-item correlations in each feature dimensions. As we can see, existing attention mechanisms still have much room to improve in modeling dynamic and diverse preferences of users.

\subsection{Position Embedding (PE)}

In the field of recommender systems, how to deal with sequence information has become a topic widely considered. For the next-item recommendation, user preferences are not fixed but change with time. The temporal recommendation of the \emph{Netflix} Prize utilized time series information on many datasets, which greatly optimized the effect of models~\cite{kang2018self}. Recent studies have also pointed out that PE can make improvements on models~\cite{huang2018csan, gehring2017convolutional}.

CNNs implicitly capture time series information by adjusting the size of the kernel and sequentially moving the kernel during training. If explicit PE is utilized at the same time, it can make some improvements to the model~\cite{nguyen2015event}. One approach of explicit PE is to take time information (such as timestamps) from datasets, and then concatenate its embedding or add its value directly to the input embedding like ATRank, but it assumes that users all have the same taste to the same time interval. Second approach of explicit PE can be a deterministic function that calculates the value at each position by sine and cosine, such as Transformer, with the same assumption mentioned above. Another approach is to generate a position matrix and add its values to input embeddings at corresponding position like CSAN, but untrainable position matrix cannot capture personalized position preferences for users and can cause an order of magnitude difference between position matrix and embeddings. In addition, PACA~\cite{cao2020position} treats each position as a position embedding, and then multiply position embedding and session embedding to capture time information of each session. But PACA's trainable embedding is only able to learn the effect of each position in the session for all the users, it cannot recognize the different effects of the same position to two users.

The PE methods mentioned above, except PACA, all generate the fixed values obtained from original datasets. However, they ignore the fact that difference between two persons even when their behavior records contain the same position information such as intervals. We call the fact as the ``personalized position information'', combine it with the ``time-aggregation'' phenomenon and name the combination as ``personalized time-aggregation'' phenomenon. Then we describe this phenomenon by personalized time position embedding and set a global trainable parameter that can automatically tune the order of magnitude of the position embedding while training.

\section{The Proposed Model}
\label{sec_proposed_model}

\subsection{Problem Formulation}

Before going into details of our proposed model, we first define the problem and basic concepts. Formally, let $\mathcal{U} = \{u_1,u_2,...,u,...\}$ and $\mathcal{I} = \{j_1,j_2,...,j,...\}$ denote the user set and item set, respectively. $\mathcal{C} = \{c_1,c_2,...,c,...\}$ is the set of item categories. For each user $u\in \mathcal{U}$, her/his sequential behaviors are represented as $ \mathcal{L}^u_t = \{\mathcal{S}^{u}_1, \mathcal{S}^{u}_2,..., \mathcal{S}^{u}_t\} $, where $t$ is the current time, and $ \mathcal{S}^{u}_i \subseteq \mathcal{I} (i\in [1,t]) $ represents the session of user $u$ at time $i$. Specifically, we divide users' behavior sequences into sessions by data, and each session represents a user's behaviors within a day. Obviously, the session $ \mathcal{S}^{u}_t $ contains user $u$'s recently purchased items which reflect her/his short-term interests at time $t$. Besides, $u$'s long-term behavior sequence at time $1\sim (t-1)$ (before time $t$), denoted by $ \mathcal{L}^{u}_{t-1} = \{\mathcal{S}^{u}_1,  \mathcal{S}^{u}_2, ...,  \mathcal{S}^{u}_{t-1}\}$, indicate $u$'s long-term preferences. We call $ \mathcal{L}^{u}_{t-1} $ and $ \mathcal{S}^{u}_t $ long-term and short-term behavior sequences $w.r.t$ time $t$, respectively. In summary, our task is to predict the target users' next behaviors and recommend appropriate items to them.

The key notations used in this paper are presented in Table \ref{tab:1}.

\begin{table}[ht]
	\caption{Key Notations and Their Descriptions}
	\centering
	\scalebox{0.8}{
		\begin{tabular}{c|l}
			\toprule[2pt]
			Notations & Explanations \\
			\midrule[1pt]
			$\mathcal{U},\mathcal{I}, \mathcal{C}$ & User, item and category ID sets\\
			$U, I, C$ & User, item and category embedding sets\\
			$\mathcal{L}^u_{t-1}, \mathcal{S}^u_t, \mathcal{L}^u_t$ & Long-term, short-term and all sequential behaviors at time $t$\\
			$L^u_{t-1}, S^u_t$ & Long-term and short-term sequential behavior embeddings at time $t$\\
			$H^u_{t-1}$ & Time-aware history representation\\
			$P^u$ & Personalized time position embedding for user $u$\\
			$W_*, b_*$ & Trainable weight matrix and bias vector\\
			$L_s$ & Long-term sequence length\\
			$d_f$ & Embedding size\\
			$\gamma$ & Trainable parameter for adjusting the order of magnitude\\
			$c^u_t$ & Dynamic user category ID at time $t$\\
			$u_{e,t}$ & User embedding at time $t$ for user $u$\\
			$u_{t-1}, u_{t}$ & User long-term and current preferences representation for user $u$\\
			$y_j$ & Label of item $j$\\
			$\lambda$ & L2-loss weight\\
			\bottomrule[2pt]
	\end{tabular}} \label{tab:1}
\end{table}

\subsection{Framework}

\begin{figure}
	\centering
	\includegraphics[scale=0.92]{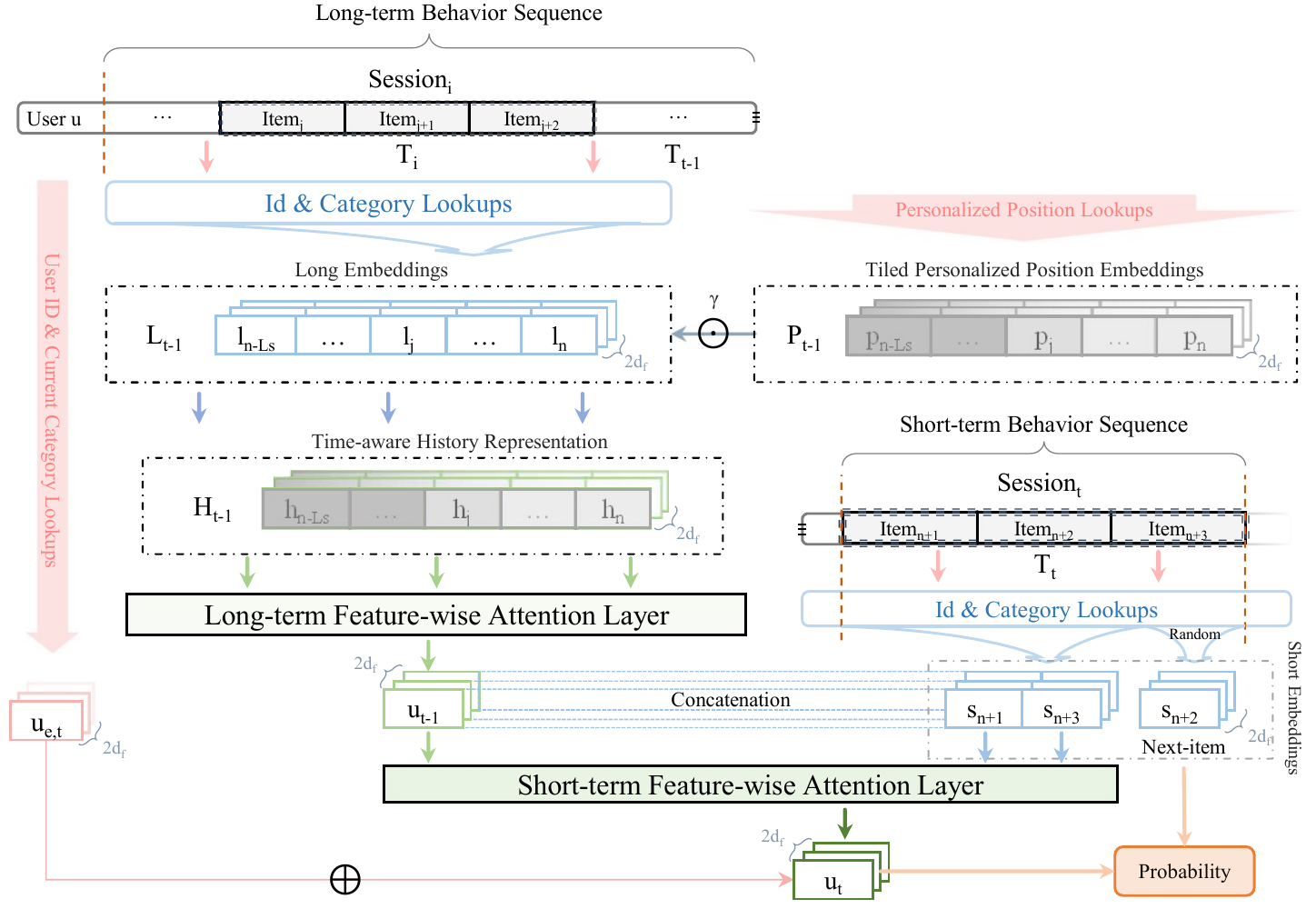}
	\caption{The framework of \tool. Specifically, we divide user's behavior records into long-term and short-term behavior records by time, and randomly select an item in short-term records as the next item to be predicted. We model user's ``personalized time-aggregation'' and capture user-specific temporal preferences by personalized time position embeddings on long-term records and get time-aware history representation. Then we can obtain user's current preferences at time $t$ by the long- and short-term feature-wise attention layers. To make the picture more concise, we have omitted some superscript $u$.}\label{fig:2}
\end{figure}

The architecture diagram of our new model is shown in Figure \ref{fig:2}. Specifically, \tool performs next item recommendation according to the following four steps: (1) \textbf{Dynamic user category extraction}: We obtain dynamic user category ID from $\mathcal{L}^u_t$ and $\mathcal{C}$, denoted by $c^u_t$. Then $c^u_t$ is utilized in lookups from $C$ to get dynamic user category embedding at time $t$. User ID embeddings $U$ and other initial embeddings are shown in Table \ref{tab:1}. Especially, we limit the long-term sequence length to $L_s$, so $L^u_{t-1}=\{l_{n-L_s},...,l_j,...,l_n\}$ where $n$ is the index of the last item in $\mathcal{L}^u_{t-1}$. Note that $l_j$ is the concatenation of item ID and category embedding, so as $s_j$ in $S^u_t$. (2) \textbf{Personalized time position embedding}: We propose trainable personalized time position embeddings, denoted by $P^u=\{p_{n-L_s},...,p_j,...,p_n\}$, to model ``personalized time-aggregation'' phenomenon in long-term sequences and capture user-specific temporal preferences. Then, we can get time-aware history representation $H^u_{t-1}=\{h_{n-L_s},...,h_j,...,h_n\}$. (3) \textbf{Long- and short-term feature-wise attention layers}: We propose long-term feature-wise attention layer to capture the long-term preferences of user $u$, denoted by $u_{t-1}$. Then we propose short-term feature-wise attention layer to combine long-term preferences $u_{t-1}$ and short-term interests from $S^u_t$ to obtain $u$'s current preferences, denoted by $u_{t}$. Besides, multi-heads integration is adopt to model the information from different semantic sub-spaces in a parallel way. (4) Finally, we utilize \tool to recommend the next item that matches the target user's current preferences $u_{t}$. 
Next, we will introduce each step in details.

\subsection{Dynamic User Category Extraction}

Most existing approaches implicitly group users, such as user-based CF~\cite{chen2020deep, koohi2016user, bellogin2012using} and social circle~\cite{purushotham2012collaborative, qian2013personalized}, which cannot explicitly model the correlations between item category. Besides, some works~\cite{papagelis2005qualitative, yang2012circle} explicitly incorporate user category information as users' feature at the cost of high time complexity. Furthermore, users' categories may change dynamically over time and it is not accurate to model user as a particular category at each time position. Especially, the item category that the user is mostly interested in at time $t$ can be considered as her/his current category to incorporate both strong user-user and user-item correlations in terms of category. Therefore, we choose the most frequent item category from $ \mathcal{L}^{u}_t $ of user $u$ as the category of user $u$ at time $t$ to obtain dynamic user category, denoted by $c^u_t$, which can be done before model training. Then we utilize lookups to extract user category embeddings from $C$ and concatenate it vertically with user ID embedding to get $u_{e,t}$. The specific implementation is as follows:
\begin{equation}
u_{e,t} = Conc(U(u), C(c^u_t)),\label{eq:1}
\end{equation}
where $u_{e,t}\in \mathbb{R}^{2d_{f}}$, $ U(.) $ and $ C(.) $ represent user ID lookups and category embedding lookups. $Conc(.)$ is the concatenation function. By replacing $U(u)$ with $I(j)$ and $c^u_t$ with $c^j$ we can also obtain the item embeddings, where $I(.)$ and $c^j$ represent item ID lookups and item category, respectively.

\subsection{Personalized time position embedding}

\begin{figure}[ht]
	\centering
	\includegraphics[scale=0.98]{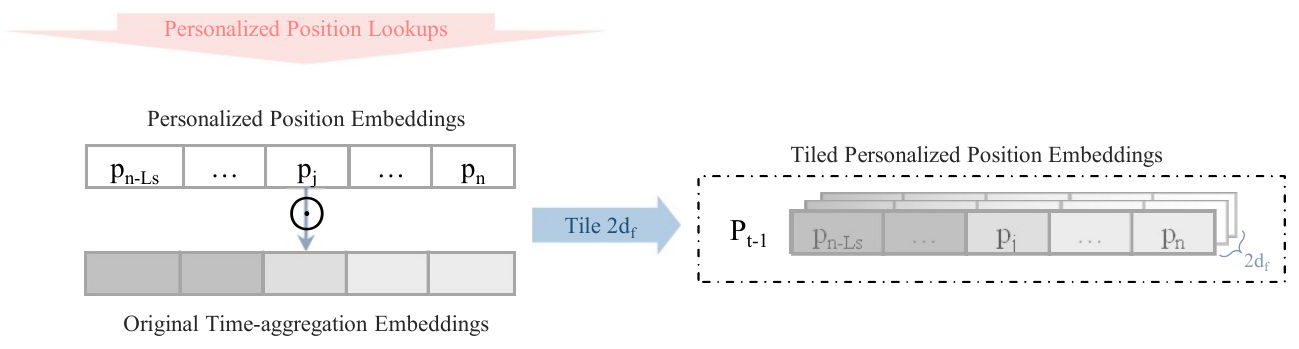}
	\caption{Original time-aggregation embedding can be obtained from the dataset, we use reciprocal value to describe time decay. The darker the color, the smaller the value. White color means no time decay. Then we multiply (mask) the personalized time position embedding to it and tile the combination to $2d_f$ dimensions.}\label{fig:3}
\end{figure}

Historical events such as promotions occurred to user (e.g. Black Friday) all happened near a certain time, which results in dense and sparse gaps in user's behavior records. We call this phenomenon ``time-aggregation'' which can implicitly reflect related environmental information, preferences information, etc. Obviously, each user has her/his own personalized behavior patterns, so different people react differently to similar context, such as the same interval of ``time-aggregation'', which is called ``personalized time-aggregation'' phenomenon. To model the ``personalized time-aggregation'' phenomenon in long-term sequences and capture user-specific temporal preferences, we propose personalized time position embedding method. During experiments, we find that the order of magnitude of position embedding is smaller than item embedding, which can weaken the influence of position embedding. Therefore, we multiply a global trainable parameter $\gamma$ (we take the initial value as $1.0$) to the position embedding. Because we use the reciprocal of the absolute value of the bucketized time difference between the past and the current time as the value of time position, the importance of the items in the current sequence starts to decrease from the last one in the time decreasing direction. At the same time, if the behaviors were generated at the same day, the values in the original time-aggregation embedding will be the same. The values in the original time-aggregation embedding can be figuratively compared to a stair, and each rung is a time position. More specifically, the personalized time position embedding method is shown in Figure \ref{fig:3}. Formally, we can obtain the time-aware historical representations as follows:
\begin{equation}
h_j = \gamma \odot p_j \odot l_j,
\end{equation}
where $\odot$ denotes the element-wise product, $h_j$ and $l_j$ are the $j$-th item in the time-aware historical representations and the corresponding long-term embeddings, respectively. $p_j$ is the $j$-th personalized time position embedding for user $u$. Then user $u$'s historical behavior records are represented as $H^u_{t-1}=\{h_{n-L_s},...,h_j,...,h_n\}\in \mathbb{R}^{4d_{f}\times L_s}$.

\subsection{Long- and Short-term Feature-wise Attention Layers}
The long- and short-term feature-wise attention layers consist of two parts: long-term feature-wise attention layer and short-term feature-wise attention layer.
\subsubsection{Long-term Feature-wise Attention Layer}

In fact, user's behavior records always contain a lot of timestamps, and earlier behavior records are generally less important than recent behavior records. Feature-wise attention is utilized on time-aware history representation $H_{t-1}$ to capture the changes of user's preferences in each dimension by the following formulas:
\begin{equation}
\begin{aligned}
att^{long}_j &= Att^{2d_{f}}(W_1, W_2, b_1, b_2, h_j) \\&= W_1^T\sigma (W_2h_j + b_2) + b_1,
\end{aligned}
\end{equation}
\begin{equation}
[a^{long}_j]_k = \frac{e^{[att^{long}_j]_k}}{\sum_{j=1}^{L_s} e^{[att^{long}_j]_k}},
\end{equation}
\begin{equation}
u_{t-1} = \sum_{j=1}^{L_s} a^{long}_j\odot h_j,
\end{equation}
where $W_1, W_2\in \mathbb{R}^{2d_{f}\times 2d_{f}}$, and $b_1, b_2\in \mathbb{R}^{2d_{f}}$ are trainable parameters. The superscript $2d_{f}$ means the dimensions of the embeddings in $Att(.)$. Especially, we choose ReLU for activation function $\sigma(.)$ to enhance non-linear capability. Now we have captured the long-term preferences of user $u$, denoted by $u_{t-1}$.

\subsubsection{Short-term Feature-wise Attention Layer}

Users' long-term preferences can be inferred from their long-term behavior records, which cannot represent their recent interests well. For the next-item recommendation, short-term interests are generally more important than users' long-term preferences, especially in sparse dataset. Therefore, we separate short-term behaviors out to emphasize the role of recent user's behavior records. However, it still requires careful consideration how to select the number of items in the short-term session. According to the ``Peak-End Rule''~\cite{do2008evaluations} in behavior economics, the most impressive and last items generally have the most significant impact on the current decision. The previous most concentrated item has been taken into account in the long-term layer, and the behaviors happened in the latest day become the focus in the short-term layer.

Besides, short-term layer combines user's short-term interests and long-term preferences together. Because the behaviors in the short-term session only happened within a day, we do not utilize time information and position embedding here to emphasize short-term interests and reduce model complexity. Formally, the short-term layer is defined as:
\begin{equation}
\begin{aligned}
att^{short}_j &= Att^{2d_{f}}(W_3, W_4, b_3, b_4, s_j) \\&= W_3^T\sigma (W_4(s_j) + b_4) + b_3,
\end{aligned}
\end{equation}
\begin{equation}
[a^{short}_j]_k = \frac{e^{[att^{short}_j]_k}}{\sum_{j=0}^{|S^{u}_{t}|+1} e^{[att^{short}_j]_k}},
\end{equation}
\begin{equation}
u_{t} = u_{e,t} \oplus \sum_{j=1}^{|S^{u}_{t}|+1} a^{short}_j\odot s_j,
\end{equation}
where $\oplus$ is an element-wise addition, $W_3, W_4\in \mathbb{R}^{2d_{f}\times 2d_{f}}$ and $b_3, b_4\in \mathbb{R}^{2d_{f}}$ are trainable parameters, $s_j\in \mathbb{R}^{2d_{f}}$, $s_j\in S^{u}_{t}$ when $j > 0$ and $s_j = u_{t-1}$ when $j = 0$. We apply $u_{e,t}$ to add context information. Similarly, we keep activation function the same as long-term attention layer.

Finally, user's current preferences are captured and represented as $u_{t}$, which leverages long-term preferences and short-term interests. Especially, we can obtain all the user's current preferences in the same way. Now with these user profiles, we can recommend the next item that is appropriate for the target user.

\subsection{Multi-heads integration}

Multi-heads method can improve parallelism and find correlations between different semantic sub-spaces~\cite{zheng2017joint, huang2018csan, zhou2018atrank, cao2020position}. Considering those advantages, we implement multi-heads on the attention mechanism mentioned above to model the information from different semantic sub-spaces and integrate them as well as their correlations in a unified and parallel way to further improve the performance of recommendation. Formally, with $m$-heads method, the attention function is defined as (superscript $^*$ means ``modified''):
\begin{equation}
^*Att^{2d_{f}} = Conc(Ahead_1,Ahead_2,...,Ahead_m),
\end{equation}
\begin{equation}
Ahead_k = Att^{2d_{f}/ m}_k(parameters).\label{eq:2}
\end{equation}
We equally separate the features of the embeddings into $Ahead_k$ ($k=1,2,...,m$), and then concatenate them in original order to avoid extra computational cost.

\subsection{Network Training}

We train \tool with the all users' behavior records in the training set, and then predict the labels(items) in the test set. Specifically, the closer the predicted label is to the truth, the more effective the model is. In this paper, we only need to predict whether the next item will be purchased or not. Therefore, we choose the unified sigmoid cross entropy loss~\cite{zhou2018atrank} for model optimization:
\begin{equation}
Loss = -\sum_{u,j} y_j\log(\sigma(f(u_t, s_j))) + (1-y_j)\log(1-\sigma(f(u_t, s_j))) + \lambda ||\Theta||^2,\label{eq:3}
\end{equation}
where $f(.)$ denotes a ranking function, which can be either a dot-product function or a more complex deep neural network, $\Theta=\{U,I,W_*,b_*\}$ and $\lambda$ is the l2-loss weight. Labels are denoted by $y\in \{0, 1\}$, and $\sigma(.)$ represents the sigmoid function. The detailed learning algorithm is presented in Algorithm~\ref{algorithm_1}.

\begin{algorithm}[ht]
	\caption{Learning Algorithm of \tool}
	\begin{algorithmic}[1]
		\Require 
		$\mathcal{L}$: long-term item behaviors,
		$\mathcal{S}$: short-term item behaviors,
		$\alpha$: learning rate,
		$d_f$: number of features,
		$\lambda$: l2-loss weight
		\Ensure 
		optimal model parameters $\Theta=\{U,I,W_*,b_*\}$ 
		\Repeat 
		\State shuffle the set of observations \{$(u,\mathcal{L}_{t-1}^u,\mathcal{S}_t^u)$\} for all users. 
		\For{each observation $(u,\mathcal{L}_{t-1}^u,\mathcal{S}_t^u)$}
		\State obtain user's current preferences $u_t$ according to Equation (\ref{eq:1})-(\ref{eq:2})
		\State compute $Loss$ according to Equation (\ref{eq:3})
		\State update $\Theta$ with gradient descent
		\EndFor
		\Until{convergence} \\
		\Return $\Theta$
	\end{algorithmic}
	\label{algorithm_1}
\end{algorithm}

\section{Experiments}
\label{sec_experiments}

In order to evaluate whether the proposed approach \tool performs well in scenarios from various fields, we utilize the datasets which cover most fields in our daily life. Specifically, the purpose of the experiments is to answer the following three questions:

\emph{RQ1}: Does \tool has better performance than other state-of-the-art models?

\emph{RQ2}: How does the parameter setting influence \tool?

\emph{RQ3}: How does each component of \tool contribute to its performance on recommendation tasks?

\subsection{Experimental Designs}
\subsubsection{Datasets}

Amazon is the world's largest e-commerce platform, which has the largest and most extensive behavior data volume. Especially, its products cover most fields in life and have good diversity. Amazon also exposes the official datasets\footnote{http://jmcauley.ucsd.edu/data/amazon/} which have filtered out users and items with less than 5 reviews and removed a large amount of invalid data. In the following experiments, only users, items, interactions, and category information are utilized. Then we perform the preprocessing according to the following two steps and the statistics for the datasets is shown in Table \ref{tab:2}.

Firstly, the users whose interactions less than 10 and the items with interactions less than 8 are removed to ensure the effectiveness of each user and item.
Secondly, we choose the users whose number of transactions is more than 4 but less than 90. This step guarantees the existence of long- and short-term behavior records and all behavior records occurred within recent three months.
\begin{table}[htbp]
	\caption{Amazon Datasets Statistics (After preprocessing)}
	\centering
	\resizebox{\linewidth}{4cm}{
		\begin{tabular}{l|ccccccc}
			\toprule[2pt]
			Datasets	& \#users & \#items & \#categories & \#samples & \makecell[c]{avg.\\items\\/category} & \makecell[c]{avg.\\behaviors\\/item} & \makecell[c]{avg.\\behaviors\\/user}\\
			\midrule[1pt]
			Electronics & 39991 & 22048 & 673 & 561100 & 32.8 & 25.4 & 14.0\\
			CDs-Vinyl & 24179 & 27602 & 310 & 470087 & 89.0 & 17.0 & 19.4\\
			Clothing-Shoes & 2010 & 1723 & 226 & 13157 & 7.6 & 7.6 & 6.5\\
			Digital-Music & 1659 & 1583 & 53 & 28852 & 29.9 & 18.2 & 17.4\\
			Office-Products & 1720 & 901 & 170 & 29387 & 5.3 & 32.6 & 17.0\\
			Movies-TV & 35896 & 28589 & 15 & 752676 & 1905.9 & 20.9 & 26.3\\
			Beauty & 3783 & 2658 & 179 & 54225 & 14.8 & 20.4 & 14.3\\
			Home-Kitchen & 11567 & 7722 & 683 & 143088 & 11.3 & 12.3 & 18.5\\
			Video-Games & 5436 & 4295 & 58 & 83748 & 74.1 & 19.5 & 15.4\\
			Toys-Games & 2677 & 2474 & 221 & 37515 & 11.2 & 15.2 & 14.0\\
			\bottomrule[2pt]	
	\end{tabular}}\label{tab:2}
\end{table}

We divide all users' behavior records into ordered sessions by day and randomly choose the item in the newest session at time $t$ as the next item while training. If the session at time $t$ contains only one item, the first item in the session at time $(t+1)$ is chosen. For the models that explicitly consider long- and short-term (SHAN, LSPM~\cite{lspm}, \tool), we consider the newest session without the chosen item as short-term session, and $1\sim (t-1)$-th sessions as long-term sessions to generate the training set. As for the other models, all sessions before time $t$ are regarded as historical sessions to get the  training set. Furthermore, we consider the item immediately after the newest session as the test item to generate the test set. 

\subsubsection{Baselines}

BPR-MF (\textbf{traditional}): Bayesian personalized ranking~\cite{rendle2009bpr} trains user's positive and negative behavior pairs to minimize the posterior probability of the difference for each user. Each item embedding is the concatenation of item ID embedding and category embedding.

CNN+Pooling (\textbf{deep neural networks; multi-head}): Max pooling operation is applied over the feature map which has kernel size 32 in the CNN structure~\cite{zheng2017joint}. We utilize this method to encode users' historical behaviors and pass all pooled features to a fully connected layer to generate user behavior embeddings.

Bi-LSTM (\textbf{deep neural networks}): LSTM~\cite{jozefowicz2015empirical} is renown by its capability of capturing implicit long- and short-term sequential data. In order to capture both forward and backward correlations among sequences, bi-directional long- and short-term memory network, called as Bi-LSTM, comes into being. In this paper, we implement this method on recommendation tasks.

ATRank (\textbf{attention; multi-head}): It considers heterogeneous users' behaviors by projecting all types of behaviors into latent semantic spaces. ATRank then utilizes self-attention layer and vanilla attention layer with DNN to obtain users' preferences~\cite{zhou2018atrank}.

PACA (\textbf{attention; position embedding}): Position-aware context attention~\cite{cao2020position} considers each time position as a trainable position vector. Then PACA captures the context of each item and the corresponding session as a session-specific feature vector by multi-layer perceptron (MLP). Attention values are generated by these two vectors.

CSAN$-$ (\textbf{attention; feature-wise; position embedding; multi-head}): Feature-wise self-attention is introduced after the embedding layer in CSAN. This model then~\cite{huang2018csan} simply adds the untrainable position encoding matrices to feature-wise self-attention. Finally, it generates user behavior embeddings through vanilla self-attention network. Since CSAN also utilizes texts, audios and images, and we do not utilize them, we represent this incomplete model in our experiments as CSAN$-$.

LSPM (\textbf{long- and short-term behaviors}): Long- and short-term preference model (LSPM)~\cite{lspm} captures each user's long-term preferences by the vector in a trainable matrix. Then it combines each user's most recent $k$ items' embeddings together to model her/his short-term preferences. Finally, LSPM combines these two preferences to obtain users' profile.

SHAN (\textbf{attention; long- and short-term behaviors}): Long- and short-term behavior records are both important, so SHAN~\cite{ying2018sequential} proposes nonlinear hierarchical attention networks to capture long- and short-term users' preferences.

\subsubsection{Evaluation Metrics}
\label{sec_eval}

We utilize three kinds of evaluation methods, area under the curve (AUC), precision and recall to evaluate models' abilities of capturing users' preferences and performing time-sensitive next-item recommendations.

AUC~\cite{zhou2018atrank} is the area enclosed by the coordinate axis under Receiver Operating Characteristic (ROC) curve. The points on ROC curve represent the value of the true tax rate (TPR) at a certain false positive rate (FPR). If the model works well, the FPR should decrease and the TPR should increase during the process of training. In other words, the better the model works, the higher the AUC value is. Formally, AUC is defined as:
\begin{equation}
AUC = \frac{1}{|U|} \sum_{u\in U} \frac{1}{|S||S'|} \sum_{s\in S} \sum_{s'\in S'} \delta(p_{u,s} > p_{u,s'}),
\end{equation}
where $S$ and $S'$ denote positive and negative sample sets, respectively. $p_{u,s}$ means the predicted probability that user $u$ may choose positive item $s$ in the test set, and $s'$ means the negative item. $\delta(.)$ is an indicator function. If $p_{u,s} > p_{u,s'}$ $\delta(p_{u,s} > p_{u,s'})$ returns 1 and 0 otherwise.

Precision@\emph{K} and Recall@\emph{K}~\cite{huang2018csan}: Precision rate refers to the ratio of the number of positive samples classified by the model to the total number of positive samples. Recall rate is the ratio of the number of positive samples classified by the model to the total number of the considered positive samples, and $K$ means only considering the $top$-$K$ items. Formally, their definitions are shown below:
\begin{equation}
Precision@K = \frac{1}{|U|} \sum_{u\in U} \frac{\sum_{s=1}^{K} fp(s, pos(u))}{K},
\end{equation}
\begin{equation}
Recall@K = \frac{1}{|U|} \sum_{u\in U} \frac{\sum_{s=1}^{K} fp(s, pos(u))}{NK(u)},
\end{equation}
where $pos(u)$ denotes the set of the ground-truth items related to user $u$, and $NK(u)$ represents the number of positive items in the $top$-$K$ predicted items of user $u$. $fp(s, pos(u))$ is an indicator which returns 1 if item $s$ is in $pos(u)$, and 0 otherwise.

AUC can measure classification ability of a classifier. For a binary classification problem, it is the ability of classifying the target values as true or false. However, AUC does not set fixed thresholds when measuring classification capabilities but considers all thresholds~\cite{dodd2003partial}. In this way, when positive and negative sample sizes are very different, the AUC value will be abnormally high, thus losing the reference value. As for next item recommendation tasks, high AUC means that the model is able to capture users' preferences, but the meaning of ``the most likely next item'' for each user is not reflected, especially in time-sensitive next-item recommendations. On the other hand, Recall@\emph{K} and Precision@\emph{K} can address this problem, so we utilize those three kinds of evaluation metrics together to make the experiments more comprehensive.

\subsubsection{Implementation Detail}

To ensure the fairness and comparability of the experiments, we keep the common parameters in each model the same and the unique parameters the optimal. For all models, we set the embedding size to 32, training batch size to 32 and testing batch size to 128. As for the specific parameters, we set the number of heads to 8 for ATRank, CNN+Pooling, CSAN$-$ and \tool, L2-loss weight to 0.01 for LSPM and to 0.00005 for other models using L2-loss. Besides, we set position kernel size to 10 for PACA, the length of recent sessions to 5 and 90 for LSPM and PACA, the length of recent long-term sessions to 10 for \tool. Moreover, the weight of short preferences is set to 1.0 for LSPM. The detailed process and analysis of choosing the critical parameters for our model \tool are shown in Section~\ref{sec_param}. Note that the parameters for other baselines are already tuned by us, but we will not show the details for them since the limited space in this paper. During the experiments, we uniformly utilize stochastic gradient descent (SGD) for training, and dynamically adjust learning rate. The initial learning rate is set to 1.0 for fast training. When each model reaches about 80\% of total training steps, learning rate is set to 0.1. All the experiments in this paper are implemented with Python 3.5 and Tensorflow 1.8.0, and run on a server with two 2.1 GHz Intel Xeon E5-2620V4 CPU, 128 GB 2133 MHz DDR4 RAM, Nvidia Tesla K40 GPU with 12 GB memory, running Ubuntu 16.04 LST. The source code of our model and the baselines with the processed datasets are publicly available\footnote{https://github.com/TsingZ0/TLSAN}.

\subsection{RQ1: Performance Analysis}
\begin{table}[ht]
	\caption{AUC on the Amazon public datasets. Bold font and underlined value indicate the optimal result and the suboptimal result, separately.}
	\centering
	\resizebox{\linewidth}{3cm}{
		\begin{tabular}{l|ccccccccc}
			\toprule[2pt]
			Datasets & ATRank & BPR-MF & CNN & CSAN-- & LSPM & PACA & Bi-LSTM & SHAN & \tool \\
			\midrule[1pt]
			Electronics & \underline{0.8659}  & 0.7457  & 0.8450  & 0.8005  & 0.7333 & 0.8322  & 0.8495  & 0.7542  & \textbf{0.9230} \\
			CDs-Vinyl & \underline{0.8999} & 0.7684 & 0.8438 & 0.7943  & 0.8594 & 0.8919 & 0.8969 & 0.7138 & \textbf{0.9651}\\
			Clothing-Shoes & 0.6761  & 0.6283  & 0.6712  & 0.5866  & 0.6443 & 0.5313  & 0.7004  & \underline{0.7284}  & \textbf{0.9363} \\
			Digital-Music & 0.8601  & 0.7896  & 0.8131  & 0.7685  & 0.8270 & \underline{0.9638}  & 0.8468  & 0.7794  & \textbf{0.9753} \\
			Office-Products & 0.9162 & 0.5610 & 0.8930 & 0.8401  & 0.7889 & 0.8994 & 0.8628 & \underline{0.9576} & \textbf{0.9773}\\
			Movies-TV & 0.8662 & 0.7654 & 0.7479 & 0.7958  & 0.8003 & 0.8055 & \underline{0.8743} & 0.7771 & \textbf{0.8986}\\
			Beauty & 0.8160  & 0.6846  & 0.7639  & 0.7620  & 0.7748 & \underline{0.9016}  & 0.8231 & 0.8953  & \textbf{0.9368} \\
			Home-Kitchen & 0.7039 & 0.6352 & 0.7075 & 0.6820  & 0.6672 & 0.8165 & 0.7373 & \underline{0.8230} & \textbf{0.8950}\\
			Video-Games & 0.8809  & 0.6609  & 0.8598  & 0.8033  & 0.8449 & 0.8763  & 0.8598  & \underline{0.9216}  & \textbf{0.9459} \\
			Toys-Games & 0.8139  & 0.6294  & 0.7788  & 0.7157  & 0.7708 & 0.8495  & 0.8012  & \underline{0.8797}  & \textbf{0.9309} \\
			\bottomrule[2pt]		
	\end{tabular}} \label{tab:3}
\end{table}

We have trained each model to converge during the experiments. The performances of them are shown in Table \ref{tab:3} and Figure \ref{fig:4}. The main observations, which are classified into traditional method, deep neural networks, attention methods, considering position embedding and considering long- and short-term behaviors, are as follows:
\begin{figure}
	\centering
	\subfigure[Recall@\emph{K}]{\includegraphics[scale=0.44]{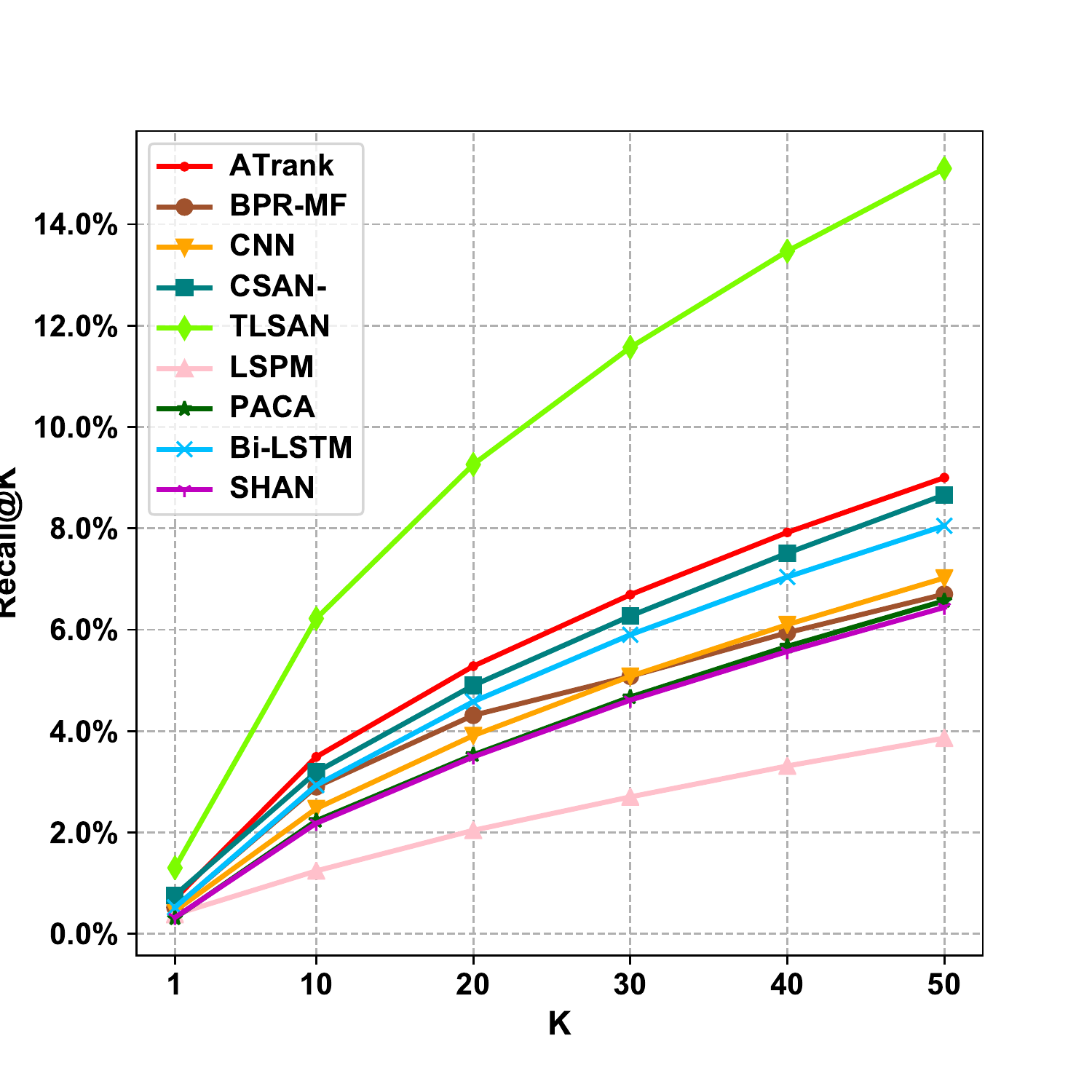}}
	\subfigure[Precision@\emph{K}]{\includegraphics[scale=0.44]{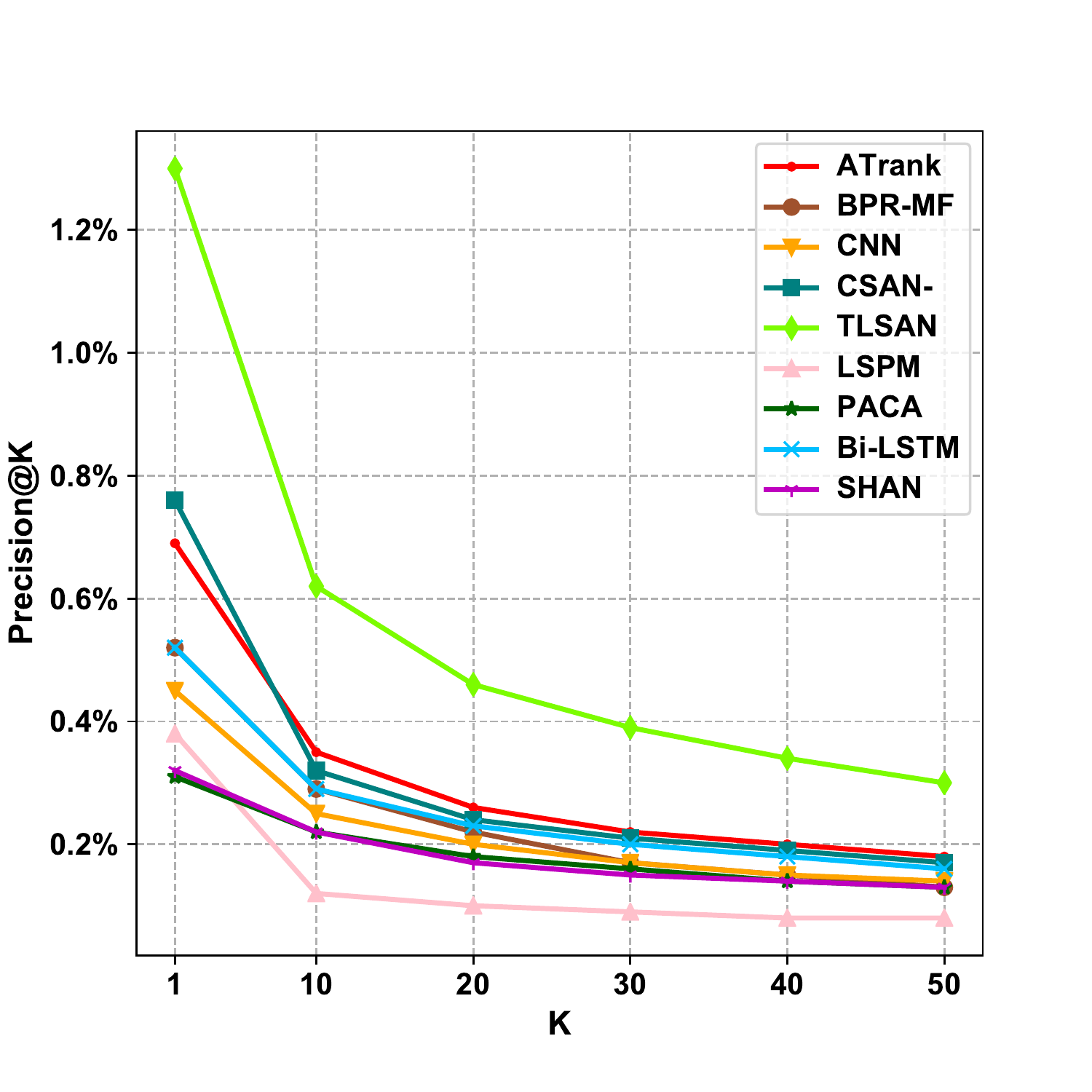}}
	\caption{Recall@\emph{K} and Precision@\emph{K} on Amazon Electronics dataset.}\label{fig:4}
\end{figure}

\begin{figure}[ht]
	\centering
	\includegraphics[scale=0.6]{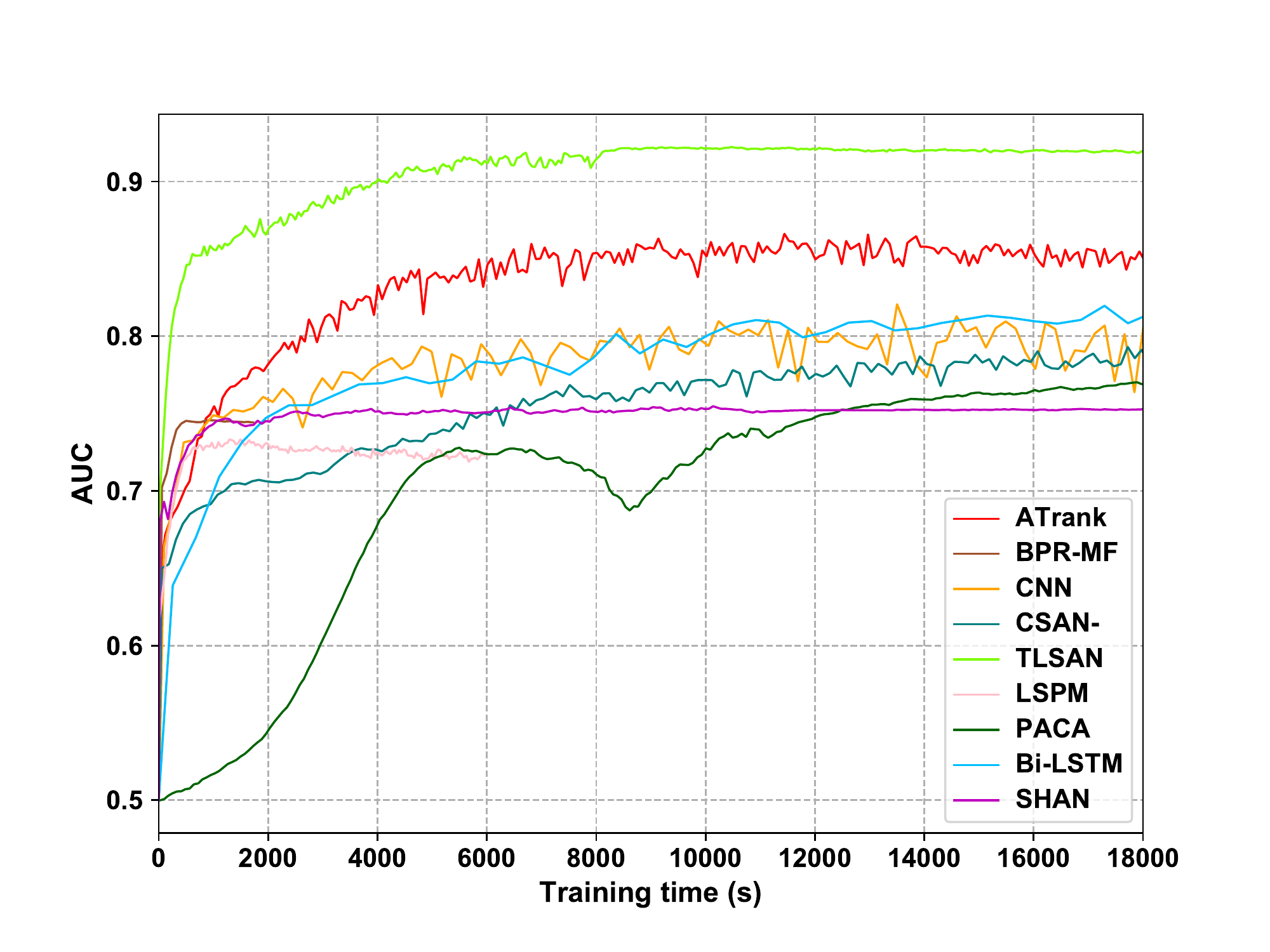}
	\caption{AUC progress on Amazon Electronics dataset. We only show the curve within 5 hours (18000 seconds) to show the different training speeds. \tool has already converged with the excellent performance, but some models (CNN, CSAN$-$, PACA, Bi-LSTM) have not converged within 5 hours.}\label{fig:5}
\end{figure}

(1) \textbf{Traditional method}: BPR-MF performs not as good as other methods on AUC, because it performs prediction based on Bayesian posterior probability, which cannot utilize temporal sequential information well. Therefore, temporal sequential information is important for recommendation. The proposed model \tool is time-aware, which inherently captures this information. 
(2) \textbf{Deep neural networks}: On average CNN is about 16.5\% better than BPR-MF for the ability of discovering users' preferences with time series information. However, its convolution kernel cannot exploit much data at one time. In the kernel, the importance of each position is the same, which ignores user's personalized preferences on position. The performance of Bi-LSTM is quite good and the AUC value on the Clothing-Shoes datasets exceeds ATRank for its excellent ability of capturing time series information. However, both CNN and Bi-LSTM have slow training speeds as shown in Figure~\ref{fig:5} for using deep neural networks. Due to the comprehensive utilization of attention mechanism, although \tool takes so many factors into consideration, it is not too complicated and can be trained faster than baselines shown as Figure~\ref{fig:5} because of parallelism. 
(3) \textbf{Attention methods}: As shown in Table~\ref{tab:3}, ATRank performs best on the Electronics dataset (second largest) except for \tool, which indicates that ATRank has better prediction ability on large datasets with attention mechanism. Specifically, ATRank performs worse than CSAN$-$ at Precision@1, and one reason is that it fails to explicitly consider ``time-aggregation''. As one of state-of-the-art models, PACA achieves good performance for attention mechanism. However, \tool performs the best among them for it explores correlations on each feature through feature-wise attention, as for attention mechanism. 
(4) \textbf{Considering position embedding}: Although PACA introduces position embedding, the embeddings in PACA are global and they are the same for all the users, which loses personalization to some extent and causes its unstable performance as shown in Table~\ref{tab:3}. CSAN$-$ still performs well on large datasets for the position matrix describes ``time-aggregation'' phenomenon. However, it does not perform as good as the proposed approach \tool. The reason is that its consistent position matrix may have different order of magnitude with historical embedding and CSAN$-$ does not incorporate user-specific temporal taste to the position embeddings, while \tool does. 
(5) \textbf{Considering long- and short-term behaviors}: LSPM and its simple version achieves good results on most datasets even on sparse dataset. The reason is that they distinguish long- and short-term preferences. However, it is not good as the proposed approach \tool. The reason is that LSPM only models short-term preferences from behavior sequences, and it regards users' embedding as the long-term preferences, which may lose both time series information and personalized position preferences. For example, LSPM gets a relatively good score on Recall@1 and Precision@1, but it becomes powerless when $K$ is larger for the reason mentioned above. The Clothing-Shoes dataset is the smallest datasets with many categories but few records (sparser) as shown in Table \ref{tab:2}, which influences most baselines' ability of capturing users' preferences. Especially, PACA cannot deal with sparse or small interaction data effectively since it does not fully exploit short-term preferences. SHAN captures long- and short-term users' preference, so it performs well on this sparse dataset as well as most remaining datasets in Table~\ref{tab:3}. \tool performs better than these baselines, since \tool explicitly considers ``personalized time-aggregation'' and captures users' long-term preferences and short-term interests when generating users' current preferences. Besides, it incorporate category-aware user-user and user-item correlations through dynamic user categories to capture more information. 

Similarly, SHAN considers both long-term and short-term behavior records, but it does not emphasize short-term interests. 
Moreover, both PACA and SHAN ignore time decay, which also influence their performance on time-sensitive next-item recommendation. Especially, Recall@\emph{K} and Precision@\emph{K} can evaluate the ability of time-sensitive next-item recommendation, while AUC is used to evaluate the ability of capturing users' preferences, as mentioned in Section~\ref{sec_eval}. So, PACA and SHAN both perform well on AUC but poorly on Recall@\emph{K} and Precision@\emph{K}. When it lacks long-term behaviors such as sparse dataset (e.g. Clothing-Shoes dataset), short-term layer in SHAN dominates and helps the model get nice performance as shown in Table \ref{tab:3}. Whether on AUC or Recall@\emph{K} and Precision@\emph{K}, \tool has improved a lot compared to other baselines especially on Recall@\emph{K} and Precision@\emph{K}, which means its time-sensitive next-item recommendation ability is excellent.

\subsection{RQ2: Parameters Analysis}
\label{sec_param}

\begin{figure}
	\centering
	\subfigure[H = 8, $L_s$ = 10]{\includegraphics[scale=0.3]{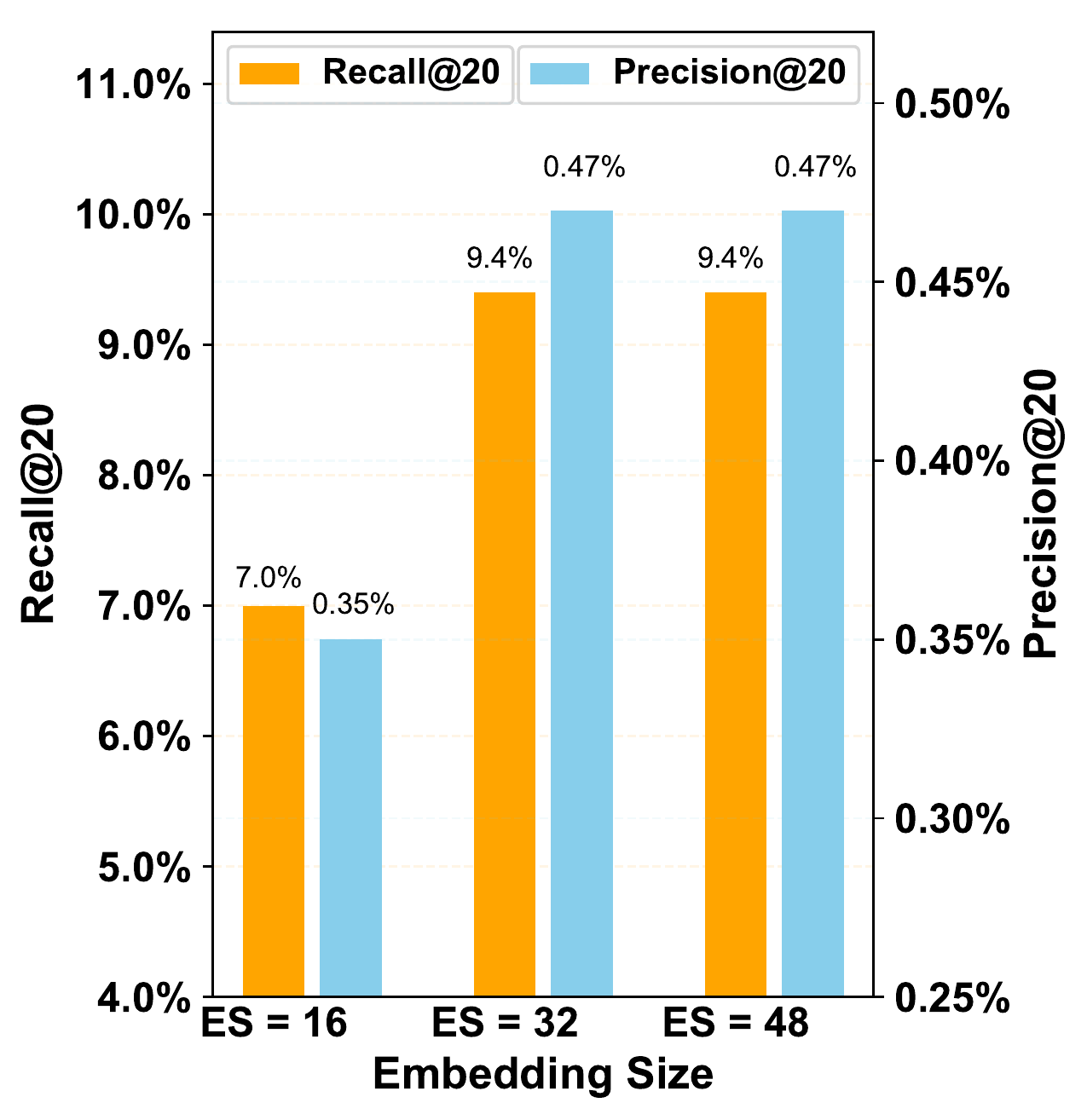}}
	\subfigure[ES = 32, $L_s$ = 10]{\includegraphics[scale=0.3]{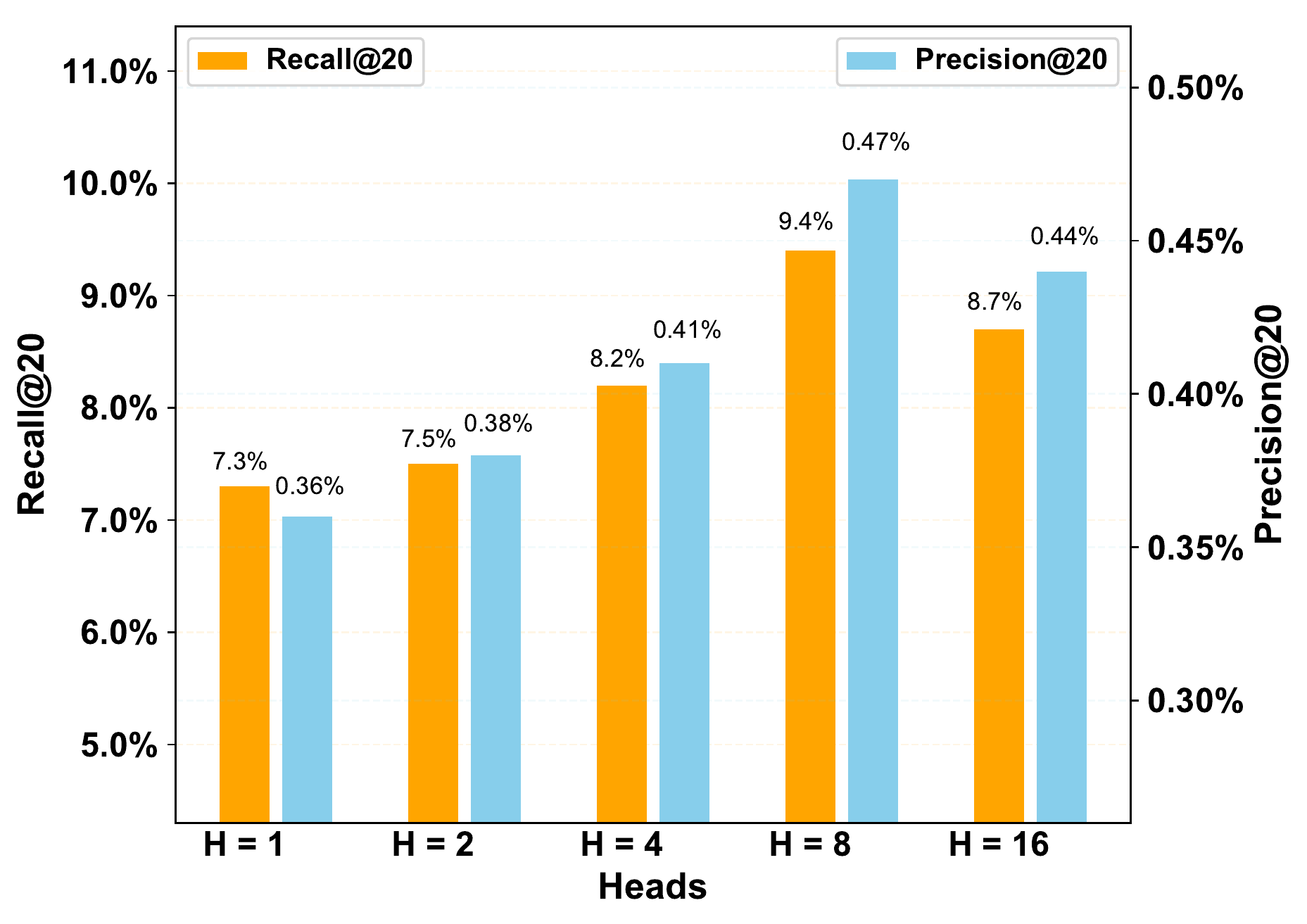}}
	\subfigure[ES = 32, H = 8]{\includegraphics[scale=0.3]{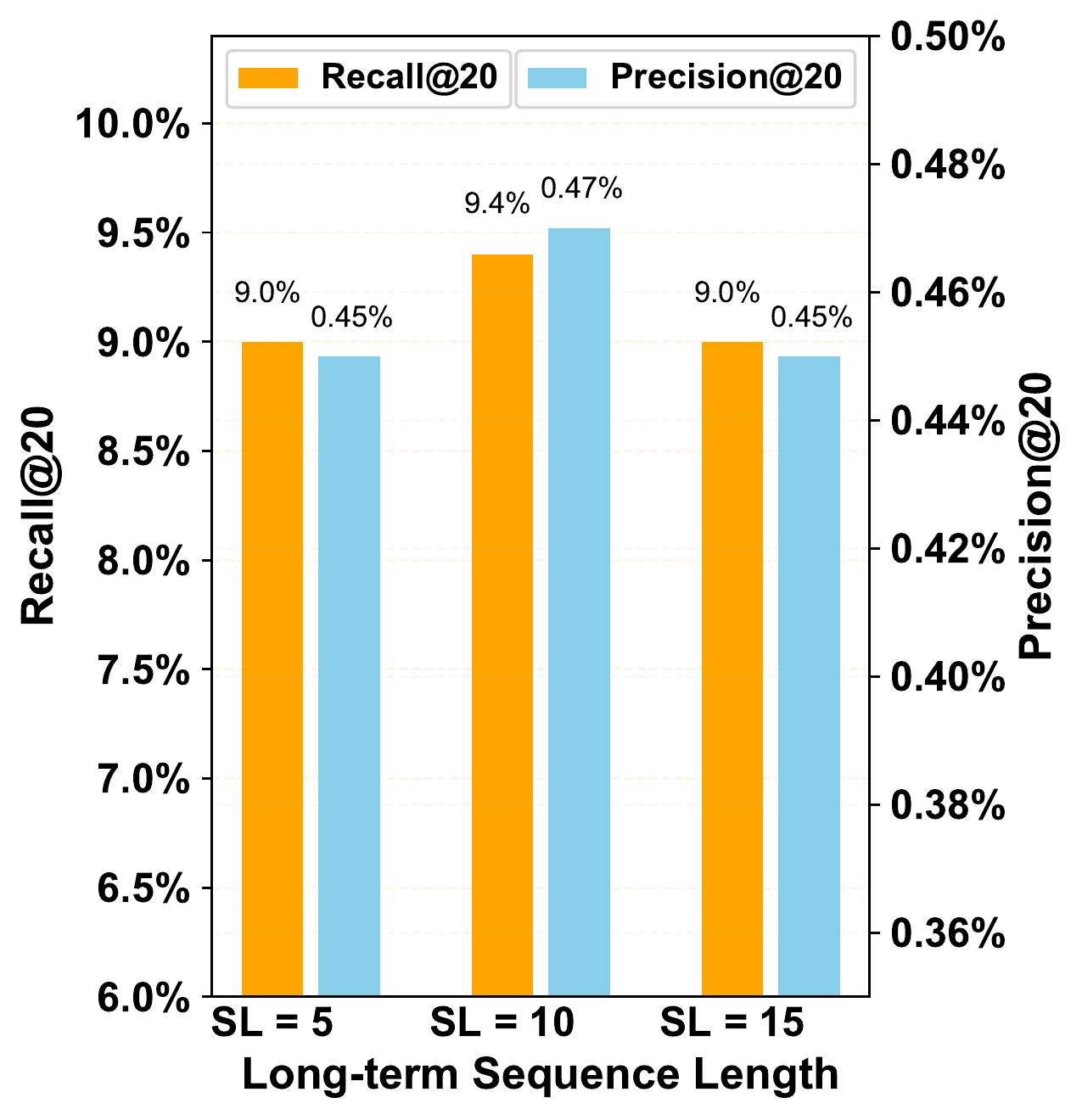}}
	\caption{Recall@20 and Precision@20 on embedding size (ES), heads (H) and long-term sequence length ($L_s$) parameters}\label{fig:6}
\end{figure}

Results of experiments for parameters analysis on the Amazon Electronics dataset are shown in Table \ref{tab:4}-\ref{tab:6} and Figure \ref{fig:6}. Specifically, we analyze the critical parameters (embedding size (ES), heads (H) and long-term sequence length ($L_s$)) as follows (Note that we only change the parameter currently being considered and keep others the same as \tool.): 

\begin{table}[ht]
	\caption{AUC on embedding size (ES).}
	\centering
	\scalebox{0.8}{
		\begin{tabular}{l|cccc}
			\toprule[2pt]
			Parameters & ES=16 & \tool & ES=48\\
			\midrule[1pt]
			AUC & 0.9129 & \underline{0.9230} & \textbf{0.9240}\\
			\bottomrule[2pt]
	\end{tabular}}\label{tab:4}
\end{table}

\begin{table}[ht]
	\caption{AUC on heads (H).}
	\centering
	\scalebox{0.8}{
		\begin{tabular}{l|cccccc}
			\toprule[2pt]
			Parameters & H=1 & H=2 & H=4 & \tool & H=16\\
			\midrule[1pt]
			AUC & 0.9199 & \underline{0.9220} & 0.9201 & \textbf{0.9230} & 0.9186\\
			\bottomrule[2pt]
	\end{tabular}}\label{tab:5}
\end{table}

\begin{table}[ht]
	\caption{AUC on long-term sequence length ($L_s$).}
	\centering
	\scalebox{0.8}{
		\begin{tabular}{l|ccccccccc}
			\toprule[2pt]
			Parameters & $L_s$=5 & \tool & $L_s$=15\\
			\midrule[1pt]
			AUC & 0.9157 & \underline{0.9230} & \textbf{0.9243}\\
			\bottomrule[2pt]
	\end{tabular}}\label{tab:6}
\end{table}

(1) Embedding Size (ES): As the embedding size increases by 16 every time, the AUC rises and plateaus when ES is 32 according to Table \ref{tab:4} and Figure \ref{fig:6} (a). The reason is that as the embedding size increases, the number of features also increases, which enable the model to capture more features or information for higher accuracy at the cost of efficiency. Besides, although the AUC on ``\tool (ES = 48)'' is higher than ``\tool (ES = 32)'', the improvement (0.108\%) is not significant compared with the improvement (1.106\%) from ``ES = 16'' to ``\tool (ES = 32)''. Simply adding features not only consumes computing power (slow down the speed), but also may force a certain feature to be split into sub-features, which may influence the performance. According to the results of Recall@20 and Precision@20 in Figure \ref{fig:6} (a), as the embedding size increases, the ability of time-sensitive next-item recommendation also reach the summit. To achieve a balance between efficieny and accuracy, ES is set to 32.

(2) Heads (H): When multi-heads method is not adopted (that is, heads = 1), \tool's performance is not bad. As the number of heads increases, the AUC reaches the optima when the heads = 8 as shown in Table \ref{tab:5} and Figure \ref{fig:6} (b). On the one hand, the multi-heads can increase parallelism and capture the interconnections between various heads. On the other hand, because the embedding size is limited to 32, the number of features of items is inversely proportional to the number of heads. Only with the proper number of heads can the two factors reach equilibrium. H = 8 achieves the best for \tool on the Amazon Electronics dataset.

(3) Long-term sequence length ($L_s$): The behaviors that happened too long ago may hardly contribute to current prediction task, and considering all the behaviors will not only weaken users' current interests but also influence efficiency. As shown in Table \ref{tab:6}, AUC keeps increasing when $L_s$ grows. The reason is that \tool can capture users' long-term preferences better with larger $L_s$, while the ability of capturing user short-term interests is already optimal in Figure \ref{fig:6} (c). Besides, \tool performs best on Recall@20 and Precision@20 when $L_s$ is 10, which means time-sensitive prediction ability has reached the summit. In the balance, we set $L_s$ to 10. Because of GPU memory saving, the training speed is also increased.

\subsection{RQ3: Component Analysis}

\begin{table}[htbp]
	\caption{AUC on \tool, NS (without short-term layer), NC (without categories), NU (without user categories), NP (without position embedding) and NG (without $\gamma$).}
	\centering
	\scalebox{0.8}{
		\begin{tabular}{l|cccccc}
			\toprule[2pt]
			components & \tool & NS & NC & NU & NP & NG\\
			\midrule[1pt]
			AUC & \textbf{0.9230} & 0.8395 & 0.8643 & 0.9225 & \underline{0.9229} & 0.9017\\
			\bottomrule[2pt]
	\end{tabular}}\label{tab:7}
\end{table}

\begin{figure}[htbp]
	\centering
	\includegraphics[scale=0.4]{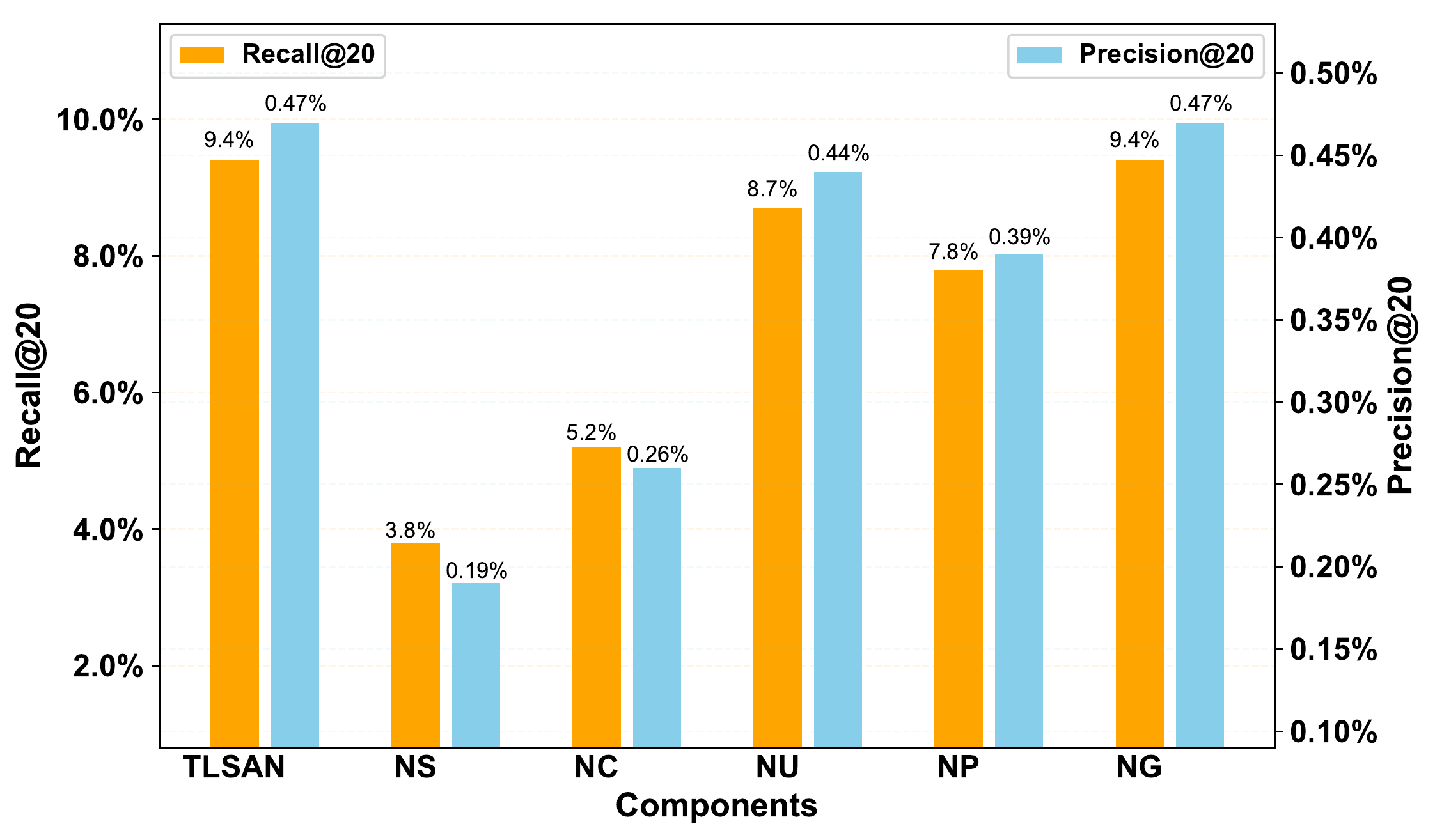}
	\caption{Recall@20 and Precision@20 on \tool, NS (without short-term layer), NC (without categories), NU (without user categories) NP (without position embedding) and NG (without $\gamma$).}\label{fig:7}
\end{figure}

The purpose of this section is to explore how each main component in \tool contributes to the recommendation performance. Specifically, the main components of \tool are short-term feature-wise attention layer, dynamic user category extraction and personalized time position embedding module. Therefore, we will remove the corresponding components to obtain the following variants and compare them with \tool: NS (without short-term layer), NC (without user and item categories), NU (without dynamic user categories), NP (without personalized time position embedding) and NG (without $\gamma$). The results on the Amazon Electronics dataset are shown in Table \ref{tab:7} and Figure \ref{fig:7}.

As can be seen from Table \ref{tab:7} and Figure \ref{fig:7}, the performance of those variants (incomplete models) is not as good as \tool,  although some of them achieve better performance than baselines. Specifically, we will explain in the following four aspects: 

(1) Many existing models consider long-term preferences of users, but they do not distinguish long-term preferences and short-term interests explicitly, so we remove short-term layer here to study its validity. As shown in Table \ref{tab:7} and Figure \ref{fig:7}, we find that the NS model with short-term layer removed is not as good as other variants. For we set the long-term sequence length limit to $L_s$, the model needs short-term sequence to provide enough information to learn both long-term preferences and short-term interests. Besides, we find that, the AUC of NS is better than some baselines such as PACA, but its performance on Recall@20 and Precision@20 is almost worse than all the baselines, which means short-term layer not only helps capture users' long-term preferences but also plays an important role in time-sensitive next-item recommendation.

(2) The comparisons between NC and NU show the effects of category attributes. As shown in in Table \ref{tab:7} and Figure \ref{fig:7}, we can observe that categories are important for improving the performance of recommendation. Due to the effectiveness of other components, the performance of NC is still better than most of the baselines and almost the same with ATRank which utilizes categories. The improvement of introducing dynamic user category is not obvious on AUC, but obvious on Recall@20 and Precision@20. Especially, the improvement on capturing users' current preferences is greater than capturing users' long-term preferences when we introduce dynamic user category.

(3) \tool has improvement on Recall@20 and Precision@20, but almost no improvement on AUC compared to NP. The reason is that, in the personalized time position embedding module, the original time-aggregation embeddings can weaken the effect of long-term behaviors and emphasize short-term interests. Owing to personalized time position embedding, the time-sensitive next-item recommendation capability of our model \tool is improved further. 

(4) NG considers personalized time position embedding, so it can capture users' short-term interests well. However, the AUC of NG is worse than NP, which means that personalized time position embedding without $\gamma$ has difference in the order of magnitude with historical embedding. We utilize a global trainable parameter $\gamma$ to automatically adjust the order of magnitude between them.

Considering the evaluation results by AUC, Recall@20 and Precision@20, we come to the following four conclusions: 1) both long and short-term preferences are important in performing accurate recommendation, and especially the latter plays a more significant role in strong time-sensitive next-item recommendations. 2) Dynamically classifying user categories according to their favorite item categories helps greatly in both capturing users' preferences and improving the recommendation performance. 3) The ``personalized time-aggregation'' effect is a phenomenon that exists in our daily life, which inspires us how to fully exploit users' behavior patterns. 4) Automatically tuning the order of magnitude between personalized time position embedding and historical embedding is helpful.

\section{Conclusion and Future Work}
\label{sec_conclusion_future}

In summary, we propose a new model: \underline{T}ime-aware \underline{L}ong- and \underline{S}hort-term \underline{A}ttention \underline{N}etwork (\tool), which recommends the next most suitable item for the target users based on their historical behavior records in the following four steps: (1) To reinforce the category-aware correlations of user-user and user-item, we consider user dynamic favorite item category as dynamic user category. (2) Then we propose a new personalized time position embedding method to describe the ``personalized time-aggregation'' phenomenon. (3) Long- and short-term feature-wise attention can generate long-term preferences and strengthened short-term interests of users to get their current preferences. (4)Finally, \tool can perform personalized sequential recommendation to target users based on their preferences. Extensive experiments are performed on Amazon datasets of different size and from different fields, and the results show that the proposed approach \tool outperforms state-of-the-art baselines. We further conduct parameter experiments to explore the process of parameter settings on \tool, and also show the effectiveness of each component through ablation experiments.

For future work, we plan to further exploit the phenomenon of ``personalized time-aggregation'' by non-linear operations for better recommendation. In addition, we will also try to incorporate more auxilary/side information such as images, audio, and comments, to further improve \tool's recommendation performance.

\section*{Acknowledgment}
This research was supported by Zhejiang Provincial Natural Science Foundation of China under No. LQ20F020015, and the Fundamental Research Funds for the Provincial University of Zhejiang under No. GK199900299012-017.



\bibliographystyle{elsarticle-num} 
\bibliography{MyReferences}


%
%
%
\end{sloppypar}
\end{document}